\documentclass[prd,twocolumn,aps, floatfix,preprintnumbers,showpacs]{revtex4}
\usepackage{amsthm}   
\usepackage{amsmath}  
\usepackage{amssymb}  
\usepackage{mathrsfs}
\usepackage[all]{xy}
\usepackage[pdftex]{graphicx}
\usepackage{indentfirst}
\usepackage[pdftex]{hyperref}

\newcommand{\be}{\begin{equation}}
\newcommand{\ee}{\end{equation}}
\newcommand{\ba}{\begin{eqnarray}}
\newcommand{\ea}{\end{eqnarray}}

\newcommand{\en}{\nonumber\\}

\newcommand{\de}{\delta}

\newcommand{\dd}[1]{\dot{#1}}

\newcommand{\cH}{\mathcal{H}}

\begin{document}
\title{Photons and Baryons before Atoms: Improving the Tight-Coupling Approximation}
\author{Francis-Yan Cyr-Racine}\email{francis@phas.ubc.ca}
\affiliation{Department of Physics and Astronomy, University of British Columbia, Vancouver, BC, V6T 1Z1, Canada}

\author{Kris Sigurdson}\email{krs@phas.ubc.ca}
\affiliation{Department of Physics and Astronomy, University of British Columbia, Vancouver, BC, V6T 1Z1, Canada}

\date{\today}
\begin{abstract}
Prior to recombination photons, electrons, and atomic nuclei rapidly scattered and behaved, almost, like a single tightly-coupled photon-baryon plasma.
We investigate here the accuracy of the tight-coupling approximation commonly used to numerically evolve the baryon and photon perturbation equations at early times.  By solving the exact perturbations equations with a stiff solver starting deep in the radiation-dominated epoch we find the level of inaccuracy introduced by resorting to the standard first-order tight-coupling approximation. We develop a new second-order approximation in the inverse Thomson opacity expansion and show that it closely tracks the full solution, at essentially no extra numerical cost.  We find the bias on estimates of cosmological parameters introduced by the first-order approximation is, for most parameters, negligible. Finally, we show that our second-order approximation can be used to reduce the time needed to compute cosmic microwave background angular spectra by as much as $\sim\!17\%$.
\end{abstract}
\maketitle

\section{Introduction}
The cosmic microwave background (CMB) radiation provides us with a picture of the Universe as it looked when the first atoms formed, about 380 000 years after the big bang. At that time, photons and baryonic matter practically ceased interacting and the Universe became transparent to radiation, allowing CMB photons to free-stream through space. To extract accurate cosmological information from CMB data it is crucial to understand the evolution of the photon-baryon plasma before decoupling.  This involves solving the Boltzmann equations for both photons and baryons coupled by a Thomson-scattering collision term  \cite{Peebles:1970ag,Wilson:1981yi,Bond:1984fp,Vittorio:1997vt,Bond:1987ub,Ma:1995ey,SugiyamaGouda(1992)}. However, the large value of the Thomson opacity ($\tau_c^{-1}$) before recombination renders these equations stiff,  and hence difficult to solve numerically. This difficulty is usually circumvented by making use of the so-called ``tight-coupling" approximation \cite{Peebles:1970ag}. In this scheme, an alternative (approximate) form of the equations is found and used to find the solution by systematically expanding the problematic terms to first order in $\tau_c$. At late times, once the Thomson opacity drops below a certain threshold, one switches back to the exact equations to determine the final answer. \\

Recently, it has been shown that uncertainties in the cosmological recombination process may lead to a bias in estimates of cosmological parameters \cite{Grin:2009ik,RubinoMartin:2009ry,Wong:2007ym}.  Could the tight-coupling approximation also result in such a bias and affect the final result of modern Boltzmann codes such as \texttt{CAMB} \cite{Lewis:1999bs} or \texttt{CMBFAST} \cite{Seljak:1996is}?  In this paper, we first investigate the accuracy of the tight-coupling approximation by directly solving the exact set of equations at all times using a stiff integration scheme.  This necessitates calculating more accurate cosmological initial conditions than has been done in the past.  While not efficient, solving the exact equations allows us to determine the level of inaccuracy introduced by resorting to the tightly-coupled limit at early times. We then design a higher-order expansion scheme and show that at second order in $k\tau_c$ and $\dot{\tau}_c$, the final solution very closely tracks that obtained by solving the exact set of equations. We are then able to compute the bias on cosmological parameter estimates introduced by resorting to the first-order tight-coupling approximation and show that it is indeed small for most cosmological parameters. Finally, and most importantly, we describe how our second-order expansion can be used to speed up the computation of CMB power spectra without loss of overall accuracy.
\section{Solution to the Exact Equations}
The first step in testing the validity of the tight-coupling approximation is to evolve the exact set of equations from early times. This requires the use of a differential equation solver able to solve stiff systems with adaptive step sizes. We utilize the \texttt{LSODA} \cite{LSODA:1983} solver which is based on the backward differentiation formula method. We find that the stiff integrator can solve the exact Boltzmann equations provided suitably accurate initial conditions are given. Indeed, the usual initial conditions for the perturbation variables used by modern Boltzmann codes are valid only in the limit of perfect coupling between photons and baryons \cite{Ma:1995ey,Bucher:1999re}. In this limit, the dipole moments of the photon and baryon distributions are exactly equal to each other and the photon quadrupole moment vanishes. However, in order to solve the exact equations at early times, one needs to initialize the relative dipole moment (usually called the slip) between the photons and baryons and the photon quadrupole moment to nonzero values.  We describe our approach to this problem in the next subsection. We then verify the convergence of the solution obtained with the stiff integrator to ensure it is stable to changes in the numerical tolerance and accuracy settings.
\subsection{Initial Conditions}\label{initial}
To find suitable initial conditions to the system of exact equations, we expand each perturbation variable in powers of $k\tau$ and $\epsilon\equiv\tau_c/\tau$
\be
\Delta(\tau,\epsilon)=\sum_{m,n}(C_{\Delta})_{mn}(k\tau)^m\epsilon^n
\ee
and substitute the result in the system of coupled differential equations (see Appendix \ref{eqns}). Here $k$ is the Fourier wave number, $\tau$ is conformal time and $\Delta(\tau,\epsilon)$ stands for any of the following perturbation variables: $\delta_c,\delta_{\gamma},\theta_{\gamma},F_{\gamma2},\delta_b, S_b\equiv\theta_b-\theta_{\gamma},\delta_{\nu},\theta_{\nu},F_{\nu2}, \eta$ (our notation closely follows that of \cite{Ma:1995ey}). We then match coefficients of like powers of $k\tau$ and $\epsilon$ to obtain a set of linear equations for the series coefficients $(C_{\Delta})_{mn}$. We then solve these linear equations to find a global series solution, demanding that the tightly-coupled solutions (adiabatic or isocurvature) are retrieved in the limit $\epsilon\rightarrow 0$. In principle, one could try to solve the full recursion relation and obtain a closed-form expression for the $(C_{\Delta})_{mn}$. In practice however, finding the first few terms of the series is sufficient to set accurate initial conditions. Using this method, we obtain the leading-order contribution to the initial value of the slip between baryons and photons for the adiabatic mode
\be
S_b(\tau)\equiv\theta_b(\tau)-\theta_{\gamma}(\tau)=\frac{\beta_1 R_b}{6(1- R_{\nu })}\omega k^4\tau^4\epsilon+\mathcal{O}(\epsilon^2),
\ee
where $\beta_l=1-l(l+2)K/k^2$ is a normalization constant, $R_{\nu}\equiv\rho_{\nu}/(\rho_{\nu}+\rho_{\gamma})$, $R_{b}\equiv\rho_{b}/\rho_{m}$ and $\omega =H_0\Omega_m/\sqrt{\Omega_r}$. The leading-order contribution of the photon quadrupole moment of the adiabatic mode is
\be
F_{\gamma2}(\tau)=\left[\frac{16}{9}+\frac{ \left(8 R_{\nu }-5\right)\omega \tau}{3\left(2 R_{\nu }+15\right) }\right]\frac{4k^2\tau^2\epsilon}{\left(4 R_{\nu }+15\right)}+\mathcal{O}(\epsilon^2).
\ee
We list the initial conditions for all of the relevant perturbation variables in Appendix \ref{init_conds}.
\subsection{Convergence of the Stiff Integration}
\begin{figure}[t]
\includegraphics[width=0.5\textwidth]{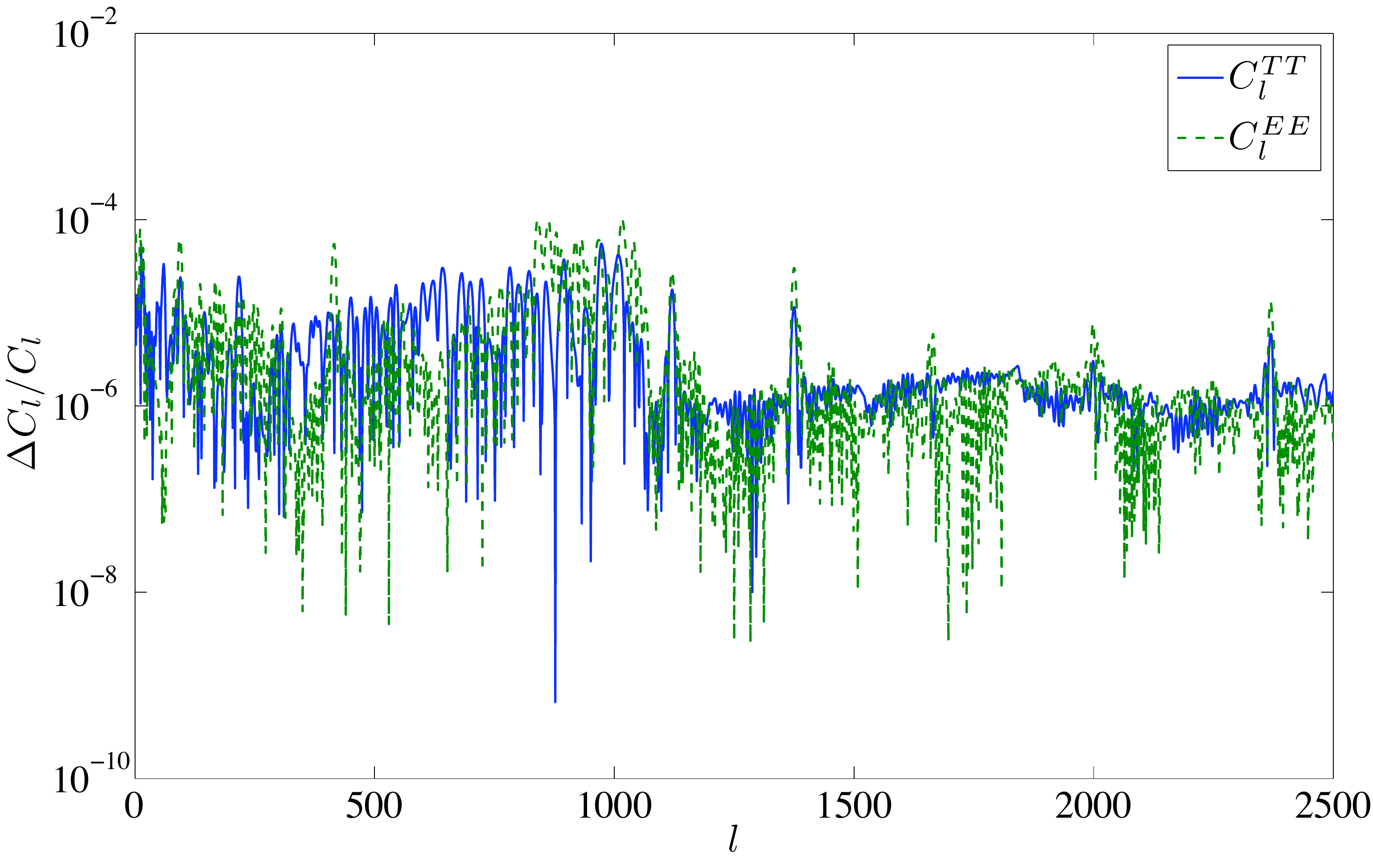}
\caption{Fractional change in $C_l^{TT}$ and $C_l^{EE}$ versus multipole moments as the three CAMB accuracy boost factors are increased from 5 to 6.  The maximum change is about $1\times10^{-4}$ for $C_l^{EE}$ and $6\times10^{-3}$ for $C_l^{TT}$.}
\label{accboost56}
\end{figure}

We verify the convergence of the stiff integrator by running several computations with increasing accuracy and comparing the resulting angular power spectra. In \texttt{CAMB}, the desired accuracy is usually selected by choosing the appropriate ``accuracy boost factors" which control, among other things, the Fourier mode sampling of the CMB anisotropy sources, the time step of the integrator, the number of multipoles kept in photon and neutrino hierarchies and the sampling of the final angular power spectrum. See Ref.~\cite{Hamann:2009yy} for a complete list.  Here, we increase the accuracy boost factors to verify for convergence but we also vary independently the tolerance of the stiff integrator to single out any error that is introduced by the solver itself. Throughout this section, we use as a benchmark model the WMAP seven-year cosmological parameter best-values \cite{Komatsu:2010fb}. 
\begin{figure}[t]
\includegraphics[width=0.5\textwidth]{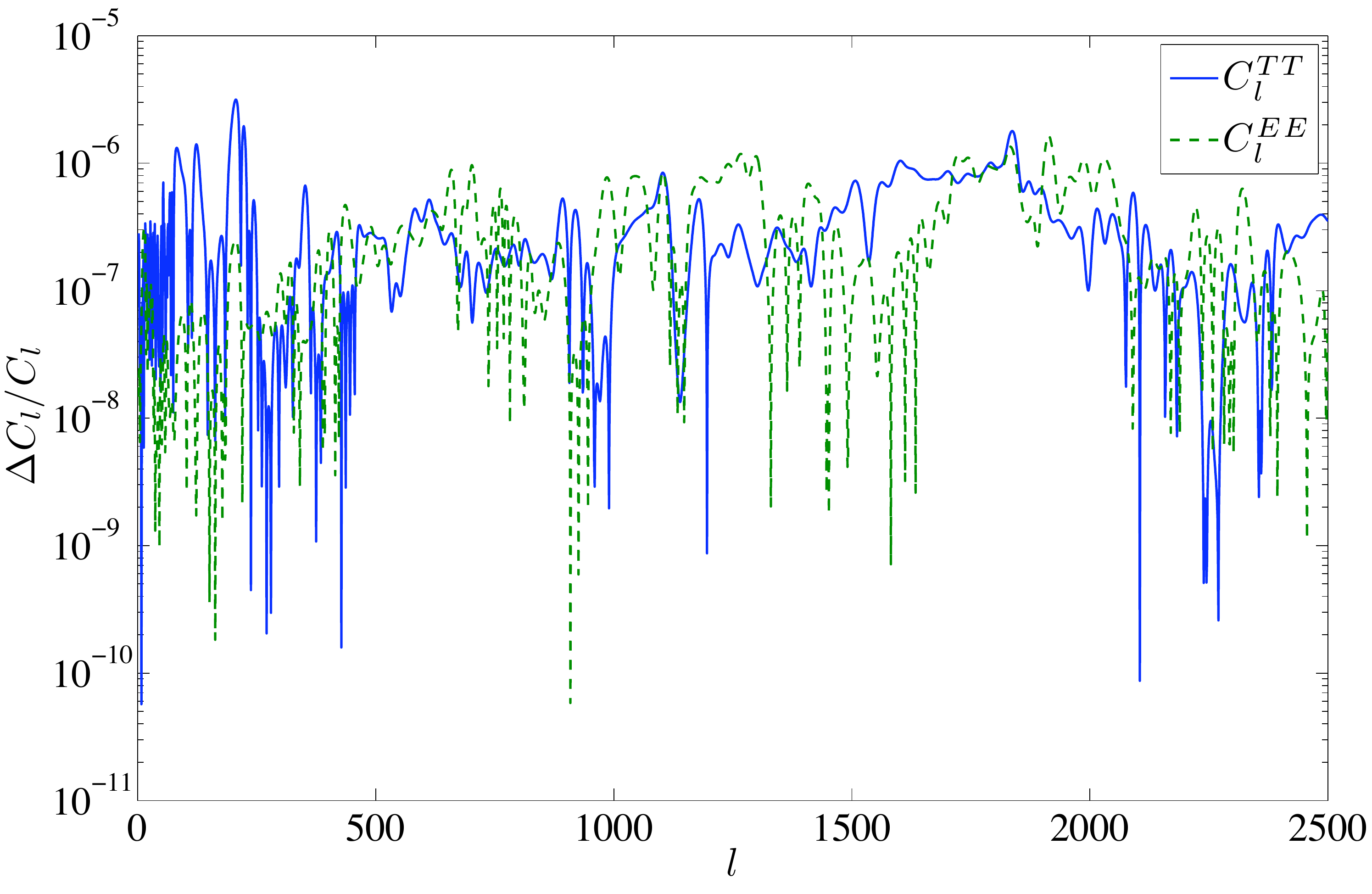}
\caption{Fractional change in $C_l^{TT}$ and $C_l^{EE}$ versus multipole moments as the relative tolerance of the stiff integrator is varied from $10^{-6}$ to $10^{-7}$. Here the CAMB accuracy boost factors are set equal to 5. The average change is $3.5\times10^{-7}$ for $C_l^{TT}$ and $3.3\times10^{-7}$ for $C_l^{EE}$.}
\label{tolerance}
\end{figure}
Figure \ref{tolerance} shows the fractional change in both $C_l^{TT}$ and $C_l^{EE}$ as a function of the multipole moment $l$ as the relative tolerance of the integrator is increased by an order of magnitude from $10^{-6}$ to $10^{-7}$. The average fractional change in the angular power spectra is approximately $3\times10^{-7}$ hence showing that the integration process has converged. Figure \ref{accboost56} shows the fractional change in both $C_l^{TT}$ and $C_l^{EE}$ as the three \texttt{CAMB} accuracy boost factors are increased from 5 to 6. We see that the $C_l$ computed with the stiff integrator have an accuracy of $0.01\%$ or better with the accuracy boost factors set to 5. We shall use this spectrum as our benchmark for testing the accuracy of our second-order tight-coupling approximation scheme which we now present in the next section.

%
\section{Second-Order Scheme}

In the usual tight-coupling approximation, the photon and baryon dipole moments are obtained by solving the two exact equations \cite{Doran:2005ep}
\ba\label{theta_gamma_tight}
\dd{\theta}_{\gamma}&=&\frac{R}{1+R}k^2\left(\frac{1}{4}\de_{\gamma}-\beta_1\frac{F_{\gamma2}}{2}\right)\en
&&+\frac{1}{1+R}\left(k^2c_s^2\de_b-\frac{\dd{a}}{a}\theta_b-\dd{S}_b\right),
\ea
\ba
\dd{\theta}_b&=&\frac{1}{1+R}\left(k^2c_s^2\de_b-\frac{\dd{a}}{a}\theta_b\right)\en
&&+\frac{R}{1+R}\left[k^2\left(\frac{1}{4}\de_{\gamma}-\beta_1\frac{F_{\gamma2}}{2}\right)+\dd{S}_b\right],
\ea
where $a$ stands for the scale factor and dots represent derivatives with respect to conformal time. The exactness of the solution to the above equations of motion depends strongly on the accuracy at which we can determine both $\dd{S}_b$ and $F_{\gamma2}$. Current CMB Boltzmann codes use a first-order expansion in $\tau_c$ to approximate the photon-baryon slip and the photon quadrupole moment. Here, we propose a method to obtain the second-order corrections in $\tau_c$ to these quantities. See \cite{Kobayashi:2007wd} for a related expansion in the context of magnetogenesis. 

\subsection{Photon-Baryon Slip}

Our starting point is the exact equation for the slip obtained from combining Eq. (\ref{theta_gamma}) and the time derivative of Eq. (\ref{slip_eq}) \cite{Doran:2005ep}:
\ba\label{slip_full}
\dd{S}_b&=&\frac{1}{1+2\frac{\dd{a}}{a}\frac{\tau_c}{1+R}}\left\{\left[\frac{\dd{\tau}_c}{\tau_c}-\frac{\dd{a}}{a}\frac{2}{1+R}\right]S_b\right.\en
&&+\frac{\tau_c}{1+R}\left[-\frac{\ddot{a}}{a}\theta_b-\ddot{S}_b
-k^2\frac{\dd{a}}{a}\left(\frac{1}{2}\de_{\gamma}-\beta_1F_{\gamma2}\right)\right.\en
&&\left.\left.+k^2\left(c_s^2\dd{\de}_b-\frac{1}{4}\dd{\de}_{\gamma}+\beta_1\frac{\dd{F}_{\gamma2}}{2}\right)\right]\right\}.
\ea
Usually, one sets $\ddot{S}_b=F_{\gamma2}=\dd{F}_{\gamma2}=0$ and neglect the prefactor on the right-hand side of Eq (\ref{slip_full}) since they contribute terms of order $\tau_c^2$ and higher. However, to obtain an equation for the photon-baryon slip valid at second order in $\tau_c$, an approximation for $\ddot{S}_b$, $F_{\gamma2}$ and $\dd{F}_{\gamma2}$ accurate to first order in $\tau_c$ is necessary. \\

The second derivative of the photon-baryon slip is computed by taking the time derivative of the right-hand side of Eq. (\ref{slip_full}). Here, we neglect terms proportional to $d^3S_b/d\tau^3$ and $\ddot{F}_{\gamma2}$. We then use the time derivative of Eqs. (\ref{de_gamma}) and (\ref{de_b}) to eliminate the second derivatives of $\de_{\gamma}$ and $\de_b$. $\ddot{h}$ is eliminated by using the i-i component of the perturbed Einstein equation
\be
\ddot{h}+2\frac{\dd{a}}{a}(2k\sigma-6\dd{\eta})-2k^2\beta_1\eta=-8\pi G a^2\sum_i3\rho_iw_i\de_i,
\ee
where $\sigma= (\dd{h}+6\dd{\eta})/2k$ is the shear. We further eliminate $\dd{\theta}_{\gamma}$ using Eq. (\ref{theta_gamma}) and set $F_{\gamma2}=\dd{F}_{\gamma2}=0$ since they contribute terms of order $\tau_c^2$ to $\ddot{S}_b$. We finally substitute the time-evolution equations for the parameters $R$, $c_s^2$, $\tau_c$ and $\dd{a}/{a}\equiv\cH$:
\ba
\dd{R}=-\cH(1-3c_s^2)R,\qquad \dd{c}_s^2=-\cH c_s^2\en
\ddot{\tau}_c = 2\dd{\cH}\tau_c+2\cH\dd{\tau}_c,\qquad \ddot{\cH}=-3\cH\dd{\cH}-\cH^3.\nonumber
\ea
Now armed with an expression for $\ddot{S}_b$, we substitute it into Eq. (\ref{slip_full}) and solve algebraically for $\dd{S}_b$. The result is given in Appendix \ref{higher}.
\subsection{Photon Quadrupole Moment}
To obtain an expression for $F_{\gamma2}$ and $\dd{F}_{\gamma2}$ accurate to second order in $\tau_c$, we use the recursion relation between higher photon multipole moments \cite{Ma:1995ey}
\be\label{recur_l}
\dd{F}_{\gamma l} = \frac{k}{2l+1}\left[lF_{\gamma (l-1)} -(l+1)\beta_lF_{\gamma (l+1)}\right]-\frac{1}{\tau_c}F_{\gamma l},
\ee
which is valid for $l\geq3$. We begin by setting $F_{\gamma5}=0$  and solve Eq. (\ref{recur_l}) with $l=4$ for $F_{\gamma4}$. We then take the derivative with respect to proper time, setting $\ddot{F}_{\gamma4}=0$. We finally solve the resulting equation for $\dd{F}_{\gamma4}$ and substitute back the result in Eq. (\ref{recur_l}). This last equation leads to an expression for $F_{\gamma4}$ valid to fourth order in $\tau_c$ (remembering that $F_{\gamma3}\propto k^2\tau_c^2$ and that $\dd{\tau}_c\propto\tau_c/\tau$):
\ba\label{Fgamma4}
F_{\gamma4}&\simeq&\frac{4}{9}k\tau_cF_{\gamma3}(1-\dd{\tau}_c)-\frac{4}{9}k\tau_c^2\dd{F}_{\gamma3}+\mathcal{O}(\tau_c^5).
\ea
Substituting the above in Eq. (\ref{recur_l}) evaluated at $l=3$ and using a similar procedure, we obtain an expression for $F_{\gamma3}$ valid to fourth order in $\tau_c$
\ba\label{Fg3}
F_{\gamma3}&\simeq&\frac{3}{7}k\tau_cF_{\gamma2}(1-\dd{\tau}_c+\dd{\tau}_c^2)-\frac{3}{7}k\tau_c^2\dd{F}_{\gamma2}(1-\dd{\tau}_c)\en
&&-\frac{16}{147}k^3\tau_c^3F_{\gamma2}+\mathcal{O}(\tau_c^5).
\ea
The last step in deriving expansions for the quadrupole moment and its derivative is to express the polarization multipole $G_{\gamma2}$ in terms of $\dd{F}_{\gamma2}$ and $F_{\gamma2}$. Similar to the above calculation, this is accomplished by using the recursion relation for the polarization multipole moments \cite{Challinor:1998xk}
\ba
\dd{G}_{\gamma l}&=&\frac{k}{2l+1}\left[lG_{\gamma(l-1)}-\frac{(l+3)(l-1)}{l+1}\beta_lG_{\gamma (l+1)}\right]\en
&&-\frac{1}{\tau_c}\left[G_{\gamma l}-\frac{2}{15}\left(\frac{3}{4}F_{\gamma2}+\frac{9}{2}G_{\gamma2}\right)\de_{l2}\right],
\ea
where $\de_{ij}$ is the Kronecker delta. Again, we set $G_{\gamma5}=0$ and follow the method outlined above to obtain
\ba\label{Gg2}
G_{\gamma2}&\simeq&\frac{F_{\gamma2}}{4}-\frac{5}{8}\tau_c\dd{F}_{\gamma2}\left(1-\frac{5}{2}\dd{\tau}_c+\frac{25}{4}\dd{\tau}_c^2\right)\en
&&-\frac{5}{56}k^2\tau_c^2F_{\gamma2}(1-6\dd{\tau}_c)+\frac{15}{27}k^2\tau_c^3\dd{F}_{\gamma2},
\ea
which is accurate to fourth order in $\tau_c$. We now have all the necessary tools to derive approximate expressions for $F_{\gamma2}$ and $\dd{F}_{\gamma2}$. We substitute Eqs. (\ref{Gg2}) and (\ref{Fg3}) in Eq. (\ref{Fg2}) and solve for $F_{\gamma2}$.  We then take the derivative with respect to conformal time and set $\ddot{F}_{\gamma2}=0$. We finally solve for $\dd{F}_{\gamma2}$ and obtain
\ba\label{dFg22}
\dd{F}_{\gamma2}&=&\frac{32}{45}\dd{\tau}_c\left( \theta_{\gamma}+k\sigma\right)\left(1-\frac{11}{6}\dd{\tau}_c\right)\en
&&+\frac{32}{45}\tau_c\left( \dd{\theta}_{\gamma}+k\dd{\sigma}\right)\left(1-\frac{11}{6}\dd{\tau}_c\right)+\mathcal{O}(\tau_c^3).
\ea
Substituting the above back in Eq. (\ref{Fg2}) we ultimately arrive at the desired expression for the photon quadrupole moment
\ba\label{Fg22}
F_{\gamma2}&=&\frac{32}{45}\tau_c\left( \theta_{\gamma}+k\sigma\right)\left(1-\frac{11}{6}\dd{\tau}_c\right)\en
&&+\frac{32}{45}\tau_c^2\left( \dd{\theta}_{\gamma}+k\dd{\sigma}\right)\left(-\frac{11}{6}\right)+\mathcal{O}(\tau_c^3).
\ea
\subsection{Computational Procedure}
As we can see from Eq. (\ref{Fg22}), our second-order expression for the photon quadrupole moment depends on $\dd{\theta}_{\gamma}$. From a practical perspective, this is problematic since it is the quantity that we are trying to determine in the first place. We overcome this difficulty by computing each quantity order by order in $\tau_c$ until the desired level of accuracy is reached.

The first step is to obtain an approximation to $F_{\gamma2}$ using Eq. (\ref{Fg22}) but keeping only the terms linear in $\tau_c$, which are independent of $\dd{\theta}_{\gamma}$. We then use this expression to calculate $\dd{\sigma}$ to first order in $\tau_c$ using the traceless space-space part of the perturbed Einstein equation
\be
k\dd{\sigma}+2\frac{\dd{a}}{a}k\sigma-k^2\eta=-8\pi G a^2\left(\rho_{\nu}F_{\nu2}+\rho_{\gamma}F_{\gamma2}\right).
\ee
Next, we calculate a zeroth-order expression for $\dd{\theta}_{\gamma}$ using Eq. (\ref{theta_gamma_tight}) with $\dd{S}_b$ and $F_{\gamma2}$ set to zero. We then use our two formulas for $\dd{\sigma}$ and $\dd{\theta}_{\gamma}$ to compute $\dd{F}_{\gamma2}$ to first order in $\tau_c$ using Eq. (\ref{dFg22}).\\

We now have all the necessary tools to calculate the photon-baryon slip to second order in $\tau_c$ using Eq. (\ref{photon_baryon_slip}). We finally use this last expression to obtain a first order approximation to $\dd{\theta}_{\gamma}$ using Eq. (\ref{theta_gamma}) which in turn is used to obtain $\dd{F}_{\gamma2}$ and $F_{\gamma2}$ accurate to second order in $\tau_c$.
\subsection{Accuracy of the Second-Order Scheme}\label{accuracy}
\begin{figure}[t]
\includegraphics[width=0.5\textwidth]{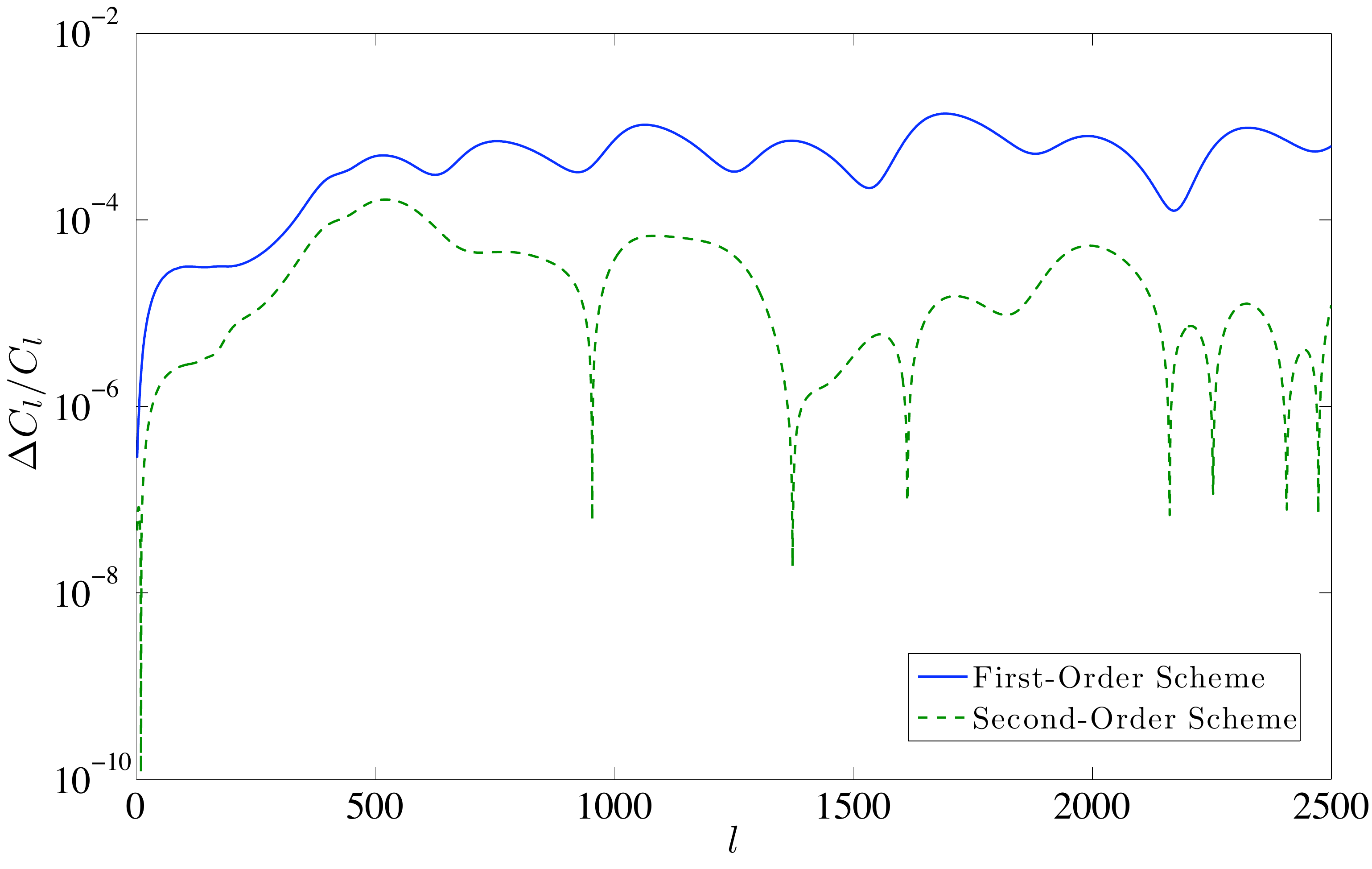}
\caption{Fractional difference of $C_l^{TT}$ between the usual first-order tight-coupling approximation and the benchmark integration (full line), and between the second-order approximation and the benchmark integration (dashed line). Here, the three sets of $C_l$s have been computed with default accuracy. The average fractional difference is $6.6\times10^{-4}$ for the first-order approximation and $5\times10^{-5}$ for the second-order approximation.}
\label{higher_vs_camb_hl}
\end{figure}

We test the accuracy of the second-order scheme by comparing the final angular power spectrum with both the stiff integrator benchmark and the usual first-order tight-coupling approximation. To isolate the effect of the second-order terms in the equations of motion, we leave untouched the algorithm that switches from the tightly-coupled equations to the exact system of equations. Improvements to the switching criteria will be discussed in Sec. \ref{comp_time}. As mentioned above, all the results presented in this section are valid for the WMAP seven-year  best-fit values for cosmological parameters.  We find that at default accuracy setting (``accuracy boost'' = 1) for all three computations, the fractional difference between the second-order scheme and the benchmark integration averaged over multipoles is about an order of magnitude smaller than the average fractional difference between the usual first-order tight-coupling approximation and the benchmark integration (see Fig. \ref{higher_vs_camb_hl}). Hence, the second-order scheme reproduces more accurately the solution to the exact equations. 

\begin{figure}[t]
\includegraphics[width=0.5\textwidth]{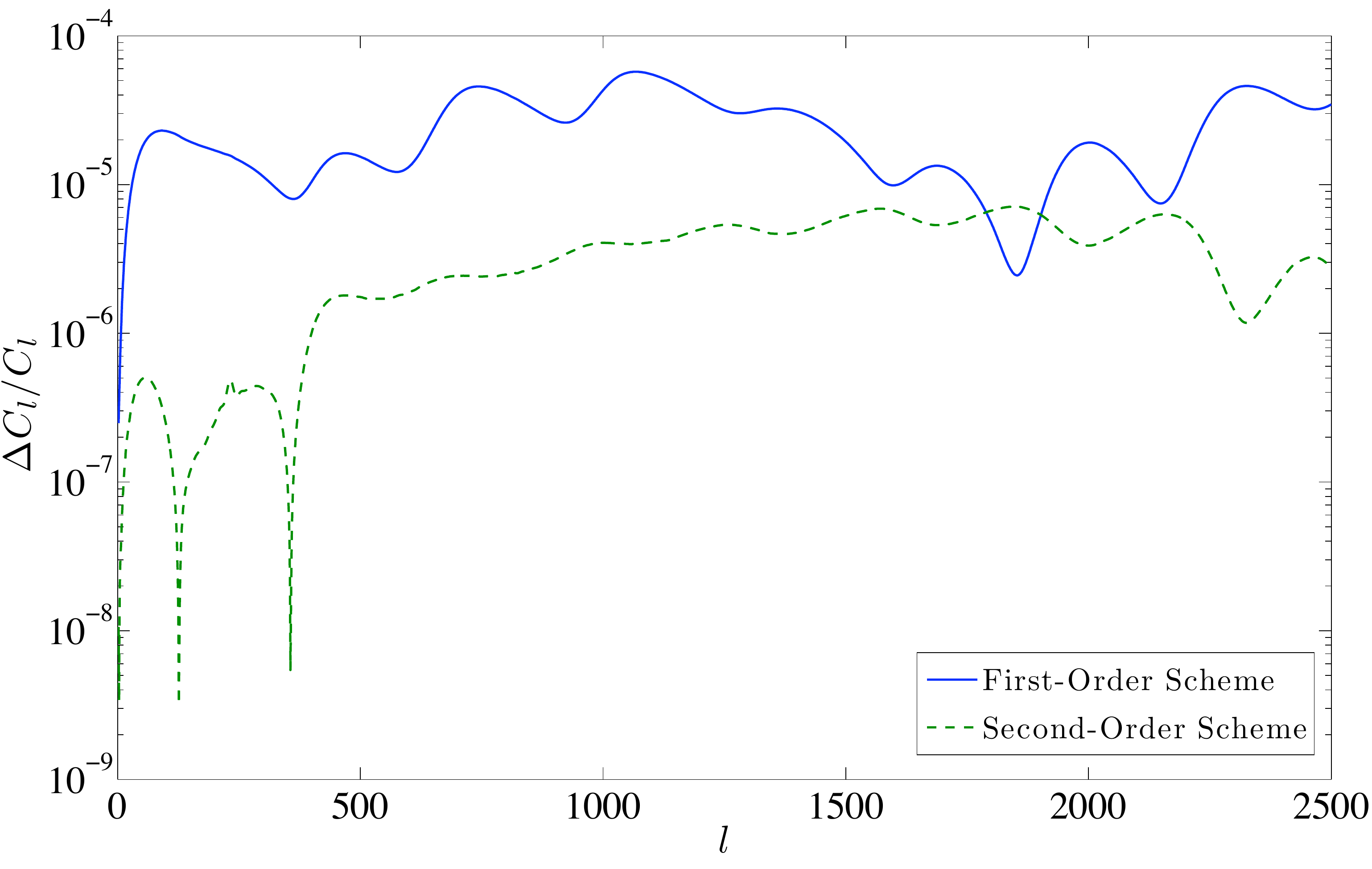}
\caption{Fractional difference of $C_l^{TT}$ between the usual first-order tight-coupling approximation and the benchmark integration (full line) and the second-order approximation and the benchmark integration (dashed line). Here, all the $C_l$s have been computed with the three accuracy boost factors set to 5. The average fractional difference is $2.4\times10^{-5}$ for the first-order approximation and $3.5\times10^{-6}$ for the second-order approximation.}
\label{higher_vs_camb_acc5}
\end{figure}

As the accuracy boost factors are increased, the second-order scheme keeps providing, on average, a more accurate answer than the first-order tight-coupling approximation. Figure \ref{higher_vs_camb_acc5} compares the angular power spectra from the two schemes with those found by integrating the exact system of equations. Although, the difference between the two codes might be insignificant for current CMB experiments, it illustrates that the next-to-leading-order code is better capturing what is happening during the tightly-coupled epoch, especially for the low multipoles. The key point however is that this better accuracy comes at almost \emph{no} additional computational cost, a point that we shall emphasize in Sec. \ref{comp_time}.

In summary, we have shown that the second-order tight-coupling approximation reproduces more closely the result found by solving the exact equations, hence showing that the tight-coupling expansion is converging toward the exact solution. For practical applications however, the percentage change in the angular power spectrum between the usual first-order approximation and the exact solution is small and well within the quoted precision from \texttt{CAMB} ($0.3\%$ at default accuracy). This implies that the first-order tight-coupling approximation should be sufficient for most practical purposes. Nevertheless, as we will describe in the next few sections, it is possible to use our second-order expansion to reduce the potential bias on cosmological parameter estimates and to speed up the code.   

We mention in passing that the precision (i.e., the size of the numerical noise) of individual $C_l$ is almost not affected by the introduction of the second-order terms in the equations of motion. Indeed, the precision of the final angular power spectrum is mainly determined by the number of k-modes evolved by the code, the number of photon multipoles that are solved for, as well as various interpolation errors. Since our new tight-coupling approximation does not modify any of the above, it is therefore natural to expect that the precision of the second-order power spectrum to remain unchanged.

\section{Bias on Cosmological Parameters}

In today's era of precision cosmology, the main purpose of CMB codes is to generate theoretical spectra that are then compared with data for cosmological parameter estimation purpose.  However, numerical errors in the theoretical spectra could lead to a slight bias on estimates of cosmological parameters \cite{Hamann:2009yy}. Since our improved tight-coupling approximation scheme leads to more accurate values of the power spectra, it is interesting to see how the bias is affected. To answer this question, we need to compute the effective $\chi^2$ between our theoretical spectra and a fiducial data set which we take to have Planck-level noise. The effective $\chi^2$ is defined by
\be
\chi^2=\sum_l(2l+1)f_{\text{sky}}\left[\text{Tr}\left(\bf{\tilde{C}}_l^{-1}\bf{\hat{C}}_l\right)+\ln\frac{|\bf{\tilde{C}}_l|}{|\bf{\hat{C}}_l|}-2\right],
\ee
where $f_{\text{sky}}$ is the observed sky fraction and ${\bf{\tilde{C}}_l}=\{C_l^{XX'}+\mathcal{N}_l^{XX'}\}$ is the theoretical covariance matrix. Here, $X$ runs over temperature (T) and polarization (E). $\bf{\hat{C}}_l$ is the data covariance matrix. If we assume that the likelihood $\mathcal{L}=\exp{(-\chi^2/2)}$  is a multivariate Gaussian and that the prior probability densities are flat, then the bias on any cosmological parameter cannot exceed $\sqrt{\chi^2}$ standard deviations. In practice however, this bound is rarely saturated. Nonetheless, a small $\chi^2$ between the data and the theory is still necessary to ensure a minimal bias. 

We generate a fiducial data set up to $l=2500$ using the method outlined in \cite{Hamann:2009yy} but with the $C_l$ obtained from the stiff solver. Again, we use the WMAP 7-year best-fit values for cosmological parameters. We take the noise to be Gaussian and isotropic with power spectrum given by
\be
\mathcal{N}_l^{XX'}=\de_{XX'}\theta_{\text{beam}}^2\Delta_X^2\exp\left[l(l+1)\frac{\theta_{\text{beam}}^2 }{8\ln2}\right],
\ee
where $\theta_{\text{beam}}$ is the beam width and $\Delta_X$ is the sensitivity per pixel. As an example, we consider the 143 GHz channel of the HFI instrument aboard Planck \cite{planck:2006uk} which has $\theta_{\text{beam}}=7.1'$, $\Delta_T=6.0 \mu K$ and $\Delta_E = 11.4 \mu K$, assuming 14 months of integrated observation. We assume a sky coverage of $f_{\text{sky}}=0.65$.

\begin{table}[b]
\begin{tabular}{|c|c|c|}
\hline
Code & $\chi^2$ & Time (s) \\
\hline\hline
\texttt{CAMB} accuracy = 1 & 2.3 & 4.8 \\
\hline
2nd Order acc. = 1 & 1.3 & 4.8 \\
\hline
\texttt{CAMB} accuracy = 2 & 0.17 &  18.2 \\
\hline
2nd Order acc. = 2 & 0.091 & 18.2\\
\hline
Opt. \texttt{CAMB} acc. = 2 & 1.1 & 15.1 \\
\hline
Opt. 2nd Order acc. = 2 &  0.10 & 15.1\\
\hline
\end{tabular}
\caption{$\chi^2$ values between fiducial Planck data and theoretical spectra gotten with the first- and second-order codes for different accuracy boost. We also give the computational time needed to generate the theoretical spectra in order to show that the greater accuracy comes at no extra numerical cost. The computational times displayed here are for a single-processor machine.}\label{chi2}
\end{table}
We list in Table \ref{chi2} the values of $\chi^2$ computed between our fiducial Planck data and our improved second-order code. For comparison, we also give the $\chi^2$ values for unmodified \texttt{CAMB} at similar accuracy boost. We see that the higher-order tight-coupling approximation leads to a better fit to the fiducial data and therefore to a smaller theoretical maximal bias on cosmological parameters at no extra numerical cost.  To estimate the real biases on cosmological parameters, we run several Markov chains using both the first- and second-order tight-coupling code together with the publicly available code \texttt{CosmoMC} \cite{Lewis:2002ah}. We restrict ourselves to the ``vanilla" 6-parameter $\Lambda$CDM model and made sure that the Gelman-Rubin ``$R-1$" convergence criteria \cite{Gelman-Rubin} is smaller than 0.005 for all the parameters under consideration.  
\begin{table}[t]
\begin{tabular}{|c|c|c|c|c|c|c|}
\hline
Code & $\Omega_bh^2$ & $\Omega_ch^2$ & $\theta$ & $\tau$ & $n_s$ & $\ln{(10^{10}A_s)}$ \\
\hline\hline
\texttt{CAMB} accuracy = 1 & 0.24 & 0.15 & 0.31 & 0.11 & 0.40 & 0.19 \\
\hline
2nd Order acc. = 1 &0.25 & 0.16 & 0.15 & 0.12 & 0.33 & 0.15 \\
\hline
\texttt{CAMB} accuracy = 2 & 0.02&0.006&0.03&0.013&0.03 &0.03\\
\hline
2nd Order acc. = 2 & 0.03 & 0.003&  0.01&0.016&0.017&0.015\\
\hline
\end{tabular} 
\caption{Biases of the 6-parameter $\Lambda$CDM model in unit of the standard deviation. We contrast the first- and second-order tight-coupling approximation and give the value of the accuracy boost factors used for each computation. }\label{bias}
\end{table}

We list in Table \ref{bias} the biases between the results from our second-order CMB code and the results from a code that used the same accuracy setting as the fiducial spectra (mimicking an error-free analysis). For comparison, we also give the biases for the usual first-order code. At default accuracy, we see that the difference between the two codes in terms of the bias is rather slim, with $\theta$ being the most dramatically affected by the second-order code. This stems from the fact that our second-order code better captures the position of the first peak as can be seen from Figs. \ref{higher_vs_camb_hl} and \ref{higher_vs_camb_acc5}.   We conclude that the numerical errors due to the first-order tight-coupling approximation does introduce a small bias to the estimate of $\theta$ at default accuracy, although it is clear that other numerical errors (k-sampling, interpolation, etc) contribute the most significant part to the biases of cosmological parameters for both codes. As the accuracy of the codes is increased, the difference in bias between the two codes becomes insignificant for parameter estimation purposes. Therefore, if one sets the accuracy of the theoretical spectra to be large enough such that the bias from numerical errors others then the first-order tight-coupling is small, then the usual tightly-coupled equations are appropriate for cosmological parameter estimation.

\section{Reducing the Computational Runtime}\label{comp_time}
Up to this point, we have used the second-order expansion in $\tau_c$ to improve the accuracy of CMB Boltzmann codes. In this section, we adopt a different point of view and take advantage of our improved tight-coupling scheme to reduce the computational time needed to evolve the perturbation equations. Indeed, the new $\mathcal{O}(\tau_c^2)$ terms in the tightly-coupled perturbation equations allow one to switch to the exact equations at a later time while keeping the same accuracy as the usual first-order expansion. Since the approximate tightly-coupled equations are easier to solve than their exact counterparts, precious computational time can be saved. Moreover, the higher accuracy of the second-order equations lets us use a larger minimal time step for the numerical integrator, hence reducing the total number of steps taken by the integrator and further cutting down the computational time. 

Our approach here is to degrade the accuracy of the second-order code by modifying the tight-coupling switching criteria, using larger time steps and cutting down the photon hierarchy until the output from this ``optimized'' code somewhat matches that of the unmodified first-order code. We then compute the $\chi^2$ value between our fiducial Planck data and the output from this optimized code and compare it to a similar calculation done with regular \texttt{CAMB}. The results are shown on the two last rows of Table \ref{chi2} where we see that we achieve a $\sim\!17\%$ computational time reduction while still retaining the accuracy of the first-order code at accuracy boost 2. 

\begin{figure}
\includegraphics[width=0.48\textwidth]{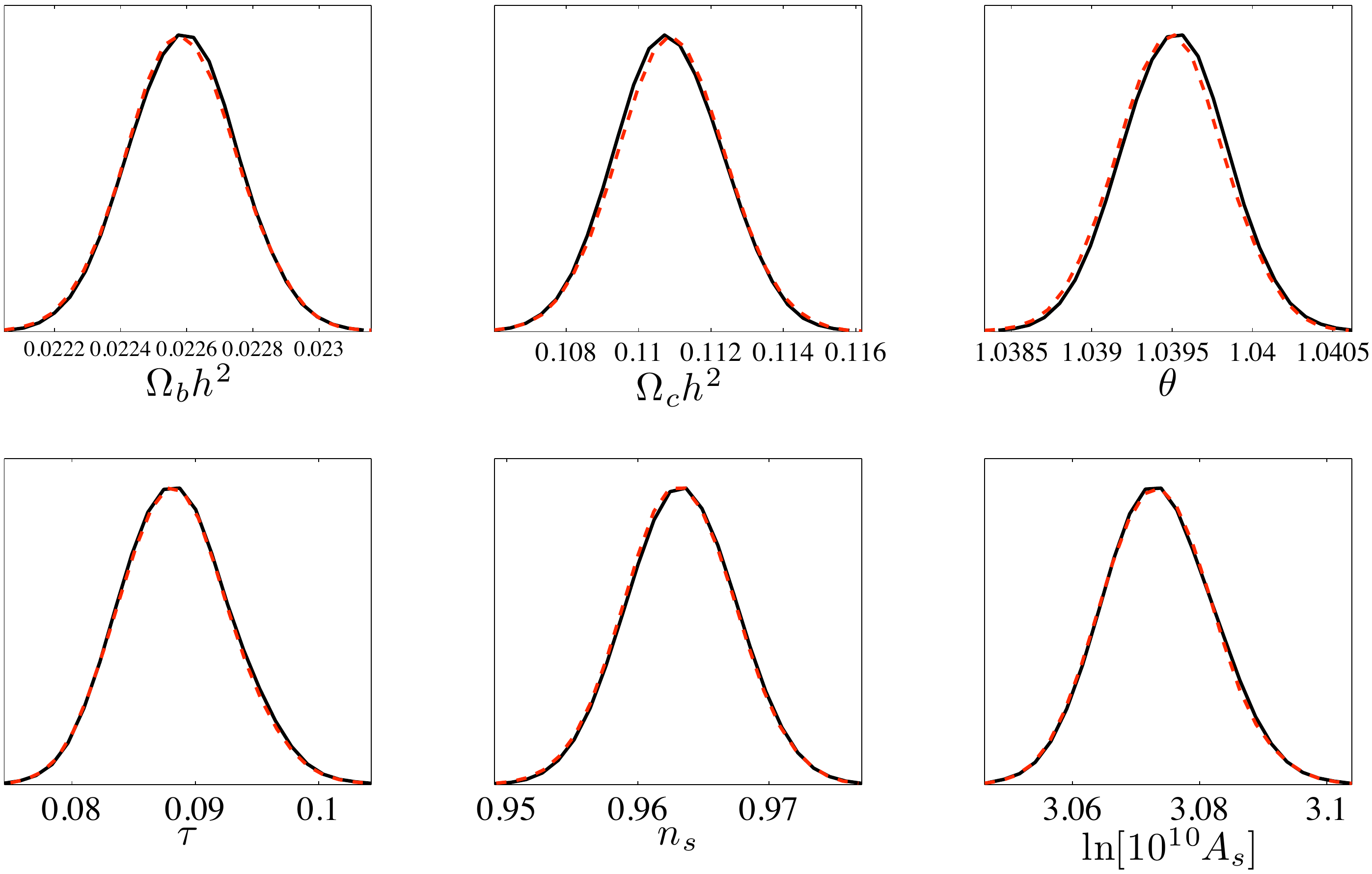}
\caption{Marginalized posterior probability distribution for the vanilla $\Lambda$CDM model. The full black line represents the result gotten using the first-order code at accuracy boost 2 while the red dotted line was obtained using our optimized second-order code.}
\label{chain_result}
\end{figure}

Although this gain in computational efficiency is modest, it can significantly reduce the amount of time necessary to run Markov chains for cosmological parameter estimation. To demonstrate this, we run 8 chains with our optimized second-order code at accuracy boost 2, generating 20000 samples per chains. We also run the similar chains with regular \texttt{CAMB} at accuracy boost 2 and make sure to have $R-1\lesssim0.005$. Figure \ref{chain_result} shows that the results for the marginalized posterior distribution are very similar between the two codes, with the distribution for $\theta$ being the most affected, although very mildly (0.09 standard deviation). However, the most important difference between the two results is that our optimized second-order code took on average $\sim16\%$ less time to complete.  Hence, our second-order tight-coupling code, in addition to leading to more accurate angular CMB spectra, can instead be used to speed up the computational time and make more efficient use of computing resources.   
%
\section{Discussion and Conclusion}
\begin{figure}
\includegraphics[width=0.5\textwidth]{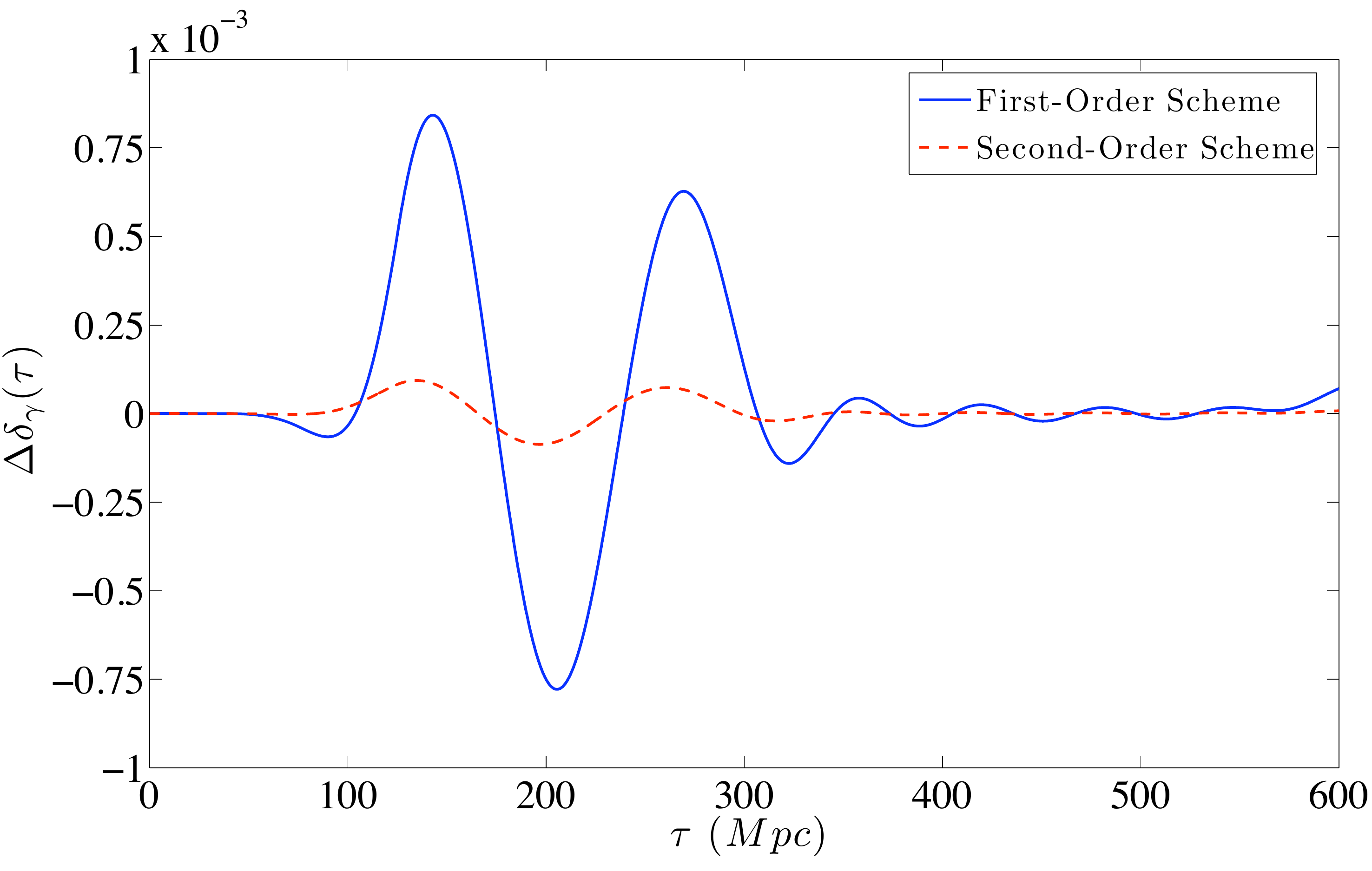}
\caption{Residuals ($\Delta\delta_{\gamma}\equiv\de_{\gamma}^{exact}-\de_{\gamma}^{approx}$) between the photon perturbation $\delta_{\gamma}$ computed using the exact equations and the solutions obtained with the first- and second-order tight-coupling approximation.}\label{delta_gamma}
\end{figure}

We have developed a second-order tight-coupling approximation to the photon-baryon perturbation equations and shown that it closely reproduces the solution to the nonapproximated equations. In practice, the main reason why our second-order tight-coupling code produces more accurate power spectra is that it better tracts the evolution of the photon perturbations. Figure \ref{delta_gamma} shows the residuals between the photon perturbation $\delta_{\gamma}$ computed using the exact equations and the solutions obtained with the first- and second-order tight-coupling approximation. We see that the second-order scheme deviates much less from the exact solution then the usual first-order scheme, leading to a more accurate value of the source term needed for the line-of-sight integration \cite{Seljak:1996is}.

In conclusion, we have investigated the accuracy of the tight-coupling approximation by solving the exact equations at all times using a stiff numerical solver. We have shown that the first-order tight-coupling approximation leads to a small accuracy lost compared to the exact solution and that this change is well within the quoted precision of the angular power spectra.  We have discussed how our second-order code has a smaller maximal possible bias on cosmological parameters than its first-order counterpart. We have shown that the bias introduced by the first-order tight-coupling is insignificant for today's cosmological experiments, unless \texttt{CAMB}'s default accuracy is used. Finally, we have shown that the improved accuracy of our second-order approximation allows one to optimize the tight-coupling switching criteria and integration parameters in order to reduce the computational time of the code.
(After this project was completed, a related work appeared \cite{Pitrou:2010ai}.)

\acknowledgments
We thank Adam Moss and Antony Lewis for useful discussions. The work of FYCR is supported by the National Science and Engineering Research Council (NSERC) of Canada.  The work of KS is supported in part by a NSERC of Canada Discovery grant.
Research infrastructure funded by  the Canada Foundation for Innovation (CFI) Leaders Opportunity Fund was essential to completing this project.

\appendix
\section{Perturbation Equations}\label{eqns}
In this Appendix, we list the perturbation equations used to solve for the initial conditions found in Appendix \ref{init_conds}. We closely follow the notation of \cite{Ma:1995ey}. Here, $\eta$ and $h$ stand for the synchronous gauge curvature perturbation variables, $a$ is the scale factor, $K$ is the inverse of the squared curvature radius, $\rho_i$ is the energy density of the $i^{th}$ specie, $w_i\equiv p_i/\rho_i$, where $p_i$ is the pressure, and a dot denotes differentiation with respect to conformal time.
\ba
k^2\beta_1\eta-\frac{1}{2}\frac{\dd{a}}{a}\dd{h}+4\pi Ga^2\sum_i\rho_i\de_i&=&0\label{const_h}\\
k^2\beta_1\dd{\eta}-\frac{K\dd{h}}{2}-4\pi Ga^2\sum_i(1+w_i)\rho_i\theta_i&=&0\\
\dd{\de}_c+\frac{1}{2}\dd{h}&=&0\\
\dd{\de}_{\nu}+\frac{4}{3}\theta_{\nu}+\frac{2}{3}\dd{h}&=&0\\
\dd{\theta}_{\nu}-\frac{k^2}{4}\left(\de_{\nu}-2\beta_1F_{\nu2}\right)&=&0\\
\dd{F}_{\nu2}-\frac{8}{15}\theta_{\nu}+\frac{3}{5}\beta_2kF_{\nu3}-\frac{4}{5}\left(\frac{\dd{h}}{3}+2\dd{\eta}\right)&=&0
\ea
\ba
\dd{\de}_{\gamma}+\frac{4}{3}\theta_{\gamma}+\frac{2}{3}\dd{h}&=&0\label{de_gamma}\\
\dd{\theta}_{\gamma}-\frac{k^2}{4}\left(\de_{\gamma}-2\beta_1F_{\gamma2}\right)-\frac{1}{\tau_c}S_b&=&0\label{theta_gamma}\\
\dd{F}_{\gamma2}-\frac{8}{15}\theta_{\gamma}+\frac{3}{5}\beta_2kF_{\gamma3}-\frac{4}{5}\left(\frac{\dd{h}}{3}+2\dd{\eta}\right)&&\en
-\frac{1}{\tau_c}\left(F_{\gamma2}-\frac{2}{15}\left(\frac{3}{4}F_{\gamma2}+\frac{9}{2}G_{\gamma2}\right)\right)&=&0\label{Fg2}\\
\dd{\de}_b+\theta_b+\frac{1}{2}\dd{h}&=&0\label{de_b}\\
S_b-\frac{\tau_c}{1+R}\left[-\frac{\dd{a}}{a}(S_b+\theta_{\gamma})-\dd{S}_b\right.&&\en
\left.+k^2\left(c_s^2\de_b-\frac{1}{4}\de_{\gamma}+\beta_1\frac{F_{\gamma2}}{2}\right)\right]&=&0\label{slip_eq}
\ea
Here, $R=(4/3)\rho_{\gamma}/{\rho_b}$. Our approach to solve these equations follow closely that of \cite{CAMB_notes}. We first use Eq. (\ref{const_h}) to eliminate $\dd{h}$ in favor of the curvature perturbation $\eta$. For simplicity, we set $c_s^2=0$. We then approximate the octupole moment of the neutrinos and photons as:
\ba
F_{\nu3}&\simeq& \frac{k\tau}{7}\left(1-\frac{4}{315}k^2\tau^2\right)F_{\nu2},
\ea
\ba
F_{\gamma3}&\simeq&\frac{3}{7}k\tau_cF_{\gamma2}.
\ea
Finally, we eliminate the photon polarization moments from (\ref{Fg2}) using
\be
G_{\gamma2}\simeq\frac{1}{4}\left(F_{\gamma2}-\frac{5\tau_c}{2}\dd{F}_{\gamma2}\right).
\ee
\section{Initial Conditions}\label{init_conds}
In this Appendix, we list the initial conditions obtained by the method outlined in Sec. \ref{initial}. Here, $R_{\nu}=\rho_{\nu}/(\rho_{\nu}+\rho_{\gamma})$, $R_b=\rho_b/\rho_m$, $\omega =H_0\Omega_m/\sqrt{\Omega_r}$, $\epsilon=\tau_c/\tau$, $S_b(\tau)\equiv\theta_b(\tau)-\theta_{\gamma}(\tau)$ is the velocity difference between baryons and photons. Note that our convention for the normalization of perturbations differs from \cite{CAMB_notes} by $\beta_1 \rightarrow -\beta_1/2$. Note also that what we label $\beta_1$ here is denoted by $\beta_2$ in \cite{CAMB_notes}. 
\begin{widetext}
\begin{enumerate}
\item Photons
\ba
\de_{\gamma}(\tau)&=&-\frac{2 \beta_1}{3}k^2\tau^2+\frac{2 \beta_1 }{15} \omega k^2\tau^3+\frac{ \beta_1 \left(4\beta_1 R_{\nu }+15\beta_1-5\right)}{27 \left(4 R_{\nu }+15\right)}k^4\tau^4-\frac{\beta_1}{24}\omega^2k^2\tau^4
\ea
\ba
\theta_{\gamma}(\tau)&=&-\frac{\beta_1}{18}k^4\tau^3-\frac{8 \beta_1}{36 R_{\nu }+135}k^4\tau^3\epsilon +\frac{\beta_1   \left(1+5 R_b-R_{\nu }\right)}{120 \left(1-R_{\nu }\right)}\omega k^4\tau^4\en
   &&-\frac{2 \beta_1   \left(2 \left(5 R_b-9\right) R_{\nu }+75 R_b+8 R_{\nu
   }^2+10\right)}{15 \left(R_{\nu }-1\right) \left(2 R_{\nu }+15\right) \left(4
   R_{\nu }+15\right)}\omega k^4\tau^4\epsilon+\frac{16 \beta_1 \left(6 R_{\nu }+181\right)}{45 \left(2 R_{\nu }+15\right) \left(4
   R_{\nu }+15\right)}k^4\tau^3\epsilon^2
\ea
\ba
F_{\gamma2}(\tau)&=&\frac{64}{9(4 R_{\nu }+15)}k^2\tau^2\epsilon+\frac{4  \left(8 R_{\nu }-5\right)}{3\left(2 R_{\nu }+15\right) \left(4 R_{\nu }+15\right)}\omega k^2\tau^3\epsilon -\frac{32  \left(6 R_{\nu }+181\right)}{9 \left(2 R_{\nu }+15\right) \left(4 R_{\nu }+15\right)}k^2\tau^2\epsilon^2\en
   &&-\frac{16   \left(2 R_{\nu } \left(12 R_{\nu }+767\right)-1855\right)}{9
    \left(2 R_{\nu }+15\right) \left(2 R_{\nu }+25\right) \left(4 R_{\nu
   }+15\right)}\omega k^2\tau^3\epsilon^2
\ea
\item Baryons
\be
\de_b(\tau)=\frac{3}{4}\de_{\gamma}(\tau)
\ee
\be
S_b(\tau)=\frac{\beta_1  R_b}{6(1- R_{\nu })}\omega k^4\tau^4\epsilon+\frac{10 \beta_1   R_b}{3 \left(1-R_{\nu }\right) \left(4 R_{\nu }+15\right)}\omega k^4\tau^4\epsilon^2-\frac{\beta_1  R_b \left(15 R_b+2 R_{\nu }-2\right)}{96  \left(R_{\nu
   }-1\right){}^2}\omega^2 k^4\tau^5\epsilon
\ee

\item Cold Dark Matter
\ba
\de_c(\tau)&=&-\frac{\beta_1}{2}k^2\tau^2+\frac{\beta_1 }{10} \omega k^2\tau^3+\frac{1}{72} \beta _1 \left(-\frac{10}{4 R_{\nu }+15}+2 \beta _1-1\right)k^4\tau^4-\frac{\beta_1}{32}\omega^2k^2\tau^4
\ea

\item Neutrinos
\ba
\de_{\nu}(\tau)&=&-\frac{2 \beta_1}{3}k^2\tau^2+\frac{2 \beta_1  }{15}\omega k^2\tau^3+\frac{1}{27} \beta _1 \left(\beta _1-\frac{1}{4 R_{\nu }+15}\right)k^4\tau^4-\frac{\beta_1}{24}\omega^2k^2\tau^4
\ea
\ba
\theta_{\nu}(\tau)&=&-\frac{\beta_1 \left(4 R_{\nu }+23\right)}{18(4 R_{\nu }+15)}k^4\tau^3+\frac{16 \beta_1 \left(1-R_{\nu }\right)}{9 \left(2 R_{\nu }+15\right) \left(4
   R_{\nu }+15\right)}k^4\tau^3\epsilon+\frac{\beta_1   \left(8 R_{\nu }^2+50 R_{\nu }+275\right)}{120 \left(2 R_{\nu}+15\right) \left(4 R_{\nu }+15\right)}\omega k^4\tau^4\en
   &&-\frac{16 \beta_1   \left(R_{\nu }-1\right) \left(2 R_{\nu }-15\right)}{15
   \left(2 R_{\nu }+15\right) \left(2 R_{\nu }+25\right) \left(4 R_{\nu
   }+15\right)} \omega  k^4\tau^4\epsilon+\frac{32 \beta_1 \left(R_{\nu }-1\right) \left(6 R_{\nu }+181\right)}{45 \left(2R_{\nu }+15\right) \left(2 R_{\nu }+25\right) \left(4 R_{\nu }+15\right)}k^4\tau^3\epsilon^2
\ea
\ba
\frac{F_{\nu2}(\tau)}{2}&=&\frac{4}{12 R_{\nu }+45}k^2\tau^2+\frac{ \left(4 R_{\nu }-5\right)}{3  \left(2 R_{\nu }+15\right)\left(4 R_{\nu }+15\right)}\omega  k^2\tau^3+\frac{64 \left(R_{\nu }-1\right)}{9 \left(2 R_{\nu }+15\right) \left(4 R_{\nu
   }+15\right)}k^2\tau^2\epsilon\en
   &&-\frac{28 \left(7 \beta _1-3\right) R_{\nu }+5 \left(175 \beta _1+27 \beta
   _2-84\right)}{189 \left(25+2R_{\nu}\right)\left(15+4R_{\nu}\right)}k^4\tau^4+\frac{ \left(4 R_{\nu } \left(2 R_{\nu }-65\right)+225\right)}{24 \left(2R_{\nu }+15\right) \left(2 R_{\nu }+25\right) \left(4 R_{\nu }+15\right)}\omega^2k^2\tau^4\en
   &&+\frac{16   \left(R_{\nu }-1\right) \left(2 R_{\nu }-15\right)}{3
   \left(2 R_{\nu }+15\right) \left(2 R_{\nu }+25\right) \left(4 R_{\nu
   }+15\right)}\omega k^2\tau^3\epsilon-\frac{32  \left(R_{\nu }-1\right) \left(6 R_{\nu }+181\right)}{9 \left(2R_{\nu }+15\right) \left(2 R_{\nu }+25\right) \left(4 R_{\nu }+15\right)}k^2\tau^2\epsilon^2
\ea
\be
F_{\nu3}(\tau)=\frac{4 }{7(12 R_{\nu }+45)}k^3\tau^3
\ee

\item Curvature (synchronous gauge)

\be
\eta(\tau)=2+\left(\frac{5}{12 R_{\nu }+45}-\frac{\beta _1}{6}\right)k^2\tau^2+\frac{80  \left(R_{\nu }-1\right)}{9 \left(2 R_{\nu }+15\right) \left(4 R_{\nu}+15\right)}k^2\tau^2\epsilon+\frac{  \left(16 \beta _1 R_{\nu }^2+20 \left(9 \beta _1+5\right) R_{\nu
   }+25 \left(18 \beta _1-5\right)\right)}{60 \left(2 R_{\nu }+15\right) \left(4 R_{\nu }+15\right)} \omega k^2\tau^3
\ee
\end{enumerate}

\section{Tight-coupling Approximation to Second Order in $\tau_c$}\label{higher}
In this Appendix, we give the key result of our improved tight-coupling approximation scheme: the photon-baryon slip to second order in $\tau_c$.

\ba\label{photon_baryon_slip}
\dd{S}_b&=&\left[\frac{\dd{\tau}_c}{\tau_c}-\cH\frac{2}{1+R}\right]S_b+\frac{\tau_c}{1+R}\left[-\frac{\ddot{a}}{a}\theta_b
-k^2\cH\left(\frac{1}{2}\de_{\gamma}-\beta_1F_{\gamma2}\right)+k^2\left(c_s^2\dd{\de}_b-\frac{1}{4}\dd{\de}_{\gamma}+\beta_1\frac{\dd{F}_{\gamma2}}{2}\right)\right]\\
&&-\left[\frac{2 R \left(3 \cH^2 c_s^2+(R+1) \dd{\cH}-3 \cH^2\right)}{(R+1)^3}\right]S_b\tau_c+\frac{\tau_c^2}{(1+R)^2}\left[\frac{\ddot{a}}{a}\frac{\cH\left( \left(2-3 c_s^2\right)R-2\right)\theta _b}{ (R+1)}+\frac{\cH k^2(1-3c_s^2) \theta _b}{3 (R+1)}\right.\en
&&+\frac{\ddot{a}}{a}\frac{k^2 c_s^2\delta _b}{(R+1)}+\frac{k^4(3c_s^2-1)  c_s^2\delta _b}{3
   (R+1)}+\frac{k^4 R (3c_s^2-1)\delta _{\gamma }}{12
   (R+1)}+\frac{\ddot{a}}{a}\frac{k^2 (2+3R) \delta
   _{\gamma }}{4 (R+1)}+\frac{\cH^2 k^2 \left(\left(2-3
   c_s^2\right)R-1\right) \delta _{\gamma }}{2 (R+1)} \en
   &&+\frac{\cH k^2 c_s^2(1+(3c_s^2-2)R) \dd{\delta}_b}{R+1}
   +\frac{\cH k^2 \left(2+\left(5-3c_s^2\right)R\right) \dd{\delta} _{\gamma }}{4 (R+1)} +\frac{2 \cH(1-3c_s^2) k^3
   \sigma }{3}+\frac{k^4 (3c_s^2-1) \beta_1\eta}{3}\en
   &&\left.+2 \cH k^2(3c_s^2-1) \dd{\eta}+\frac{k^2(1-3c_s^2) \Delta}{6}\right]+\left[\frac{4 \frac{\ddot{a}}{a} \theta _b-4  k^2 c_s^2 \dd{\delta} _b+2  \cH k^2
   \delta _{\gamma }+ k^2\dd{\delta} _{\gamma }}{2  (R+1)^2}\right]\tau_c\dd{\tau}_c-\frac{4 \cH R}{(R+1)^2}\dd{\tau}_cS_b+\mathcal{O}(\tau_c^3)\nonumber
 \ea
 Here, $\Delta=8\pi Ga^2(\rho_{\gamma}\de_{\gamma}+\rho_{\nu}\de_{\nu}+3c_s^2\rho_b\de_b)$.

\end{widetext}


\end{document}